\documentclass[journal,draftcls,onecolumn,12pt,twoside]{IEEEtranTCOM}
\normalsize
\usepackage[cmex10]{amsmath}
\usepackage{amssymb}
\usepackage[pdftex]{graphicx}
\usepackage{subcaption}

\begin{document}
	
\title{Accumulative Iterative Codes Based on Feedback}

\author{Alberto~G.~Perotti,~\IEEEmembership{Senior Member,~IEEE,}
	 Branislav~M.~Popovi\'c, and
	Anahid~R.~Safavi,~\IEEEmembership{Member,~IEEE}
\thanks{The authors are with Huawei Technologies Sweden AB,
Skalholtsgatan 9-11, SE--164 94 Kista, Sweden,
e-mail: \texttt{\small\{alberto.perotti, branislav.popovic, anahid.r.safavi\}@huawei.com}}}

\markboth{IEEE Transactions on Communications}%
{Submitted paper}

\maketitle

\begin{abstract}
The Accumulative Iterative Code (AIC) proposed in this work is a new error
correcting code for channels with feedback.
AIC sends the information message to the receiver in a number of 
transmissions, where the initial transmission contains the uncoded
message and each subsequent transmission informs the receiver about 
the locations of the errors that corrupted the previous transmission.
Error locations are determined based on the forward
channel output, which is made available to the transmitter through
the feedback channel.

AIC achieves arbitrarily low error rates, thereby being suitable
for applications demanding extremely high reliability.
In the same time, AIC achieves spectral efficiencies very close to
the channel capacity in a wide range of signal-to-noise ratios even
for transmission of short information messages.
\end{abstract}



\section{Introduction}\label{sec:intro}
\IEEEPARstart{A}{chieving} wired-like communication performances
through wireless connections is an extremely ambitious goal. 
Nonetheless, some of the most advanced applications of future mobile
cellular networks require levels of reliability and latency similar to
their wired counterparts.
In the industrial automation domain~\cite{bib:3GPPtr22-804}, for example, there
is increasing interest for providing wireless connectivity to devices involved
in the assistance and supervision of production processes, and even in real-time
motion control of production machinery.
In real-time motion control, the controller application and the controlled
machinery are connected through radio links.
All the messages exchanged over the radio links have to be properly secured
and the probability of two consecutive packet errors must be made negligible.
According to~\cite{bib:3GPPtr22-804}, a single packet error may be tolerable,
whereas two consecutive packet errors may cause damages to the controlled
machines and cause interruptions to the production processes.

In conventional transmission systems, high reliability
is obtained by Error Correction Coding (ECC).
Conventional ECC methods already provide reliable transmission with
Spectral Efficiency (SE) very close to the channel capacity.
However, the only way for obtaining reliable transmission and
capacity-approaching SE at the same time is by transmission of long codewords.
With short codes, the achievable SE is significantly smaller than the
channel capacity, as predicted by the accurate analytical
characterization in~\cite{bib:PPV10}.
In practice, even the best error correction codes, such as
the 3GPP New Radio (NR) Low-Density Parity Check (LDPC) codes and
\emph{polar codes}~\cite{bib:NRr15-212},
are unable to provide capacity-approaching SE and high reliability 
at the same time~\cite{bib:VucComMag19}.
The gap between channel capacity and SE of short-codeword transmission is
remarkably large for low signal-to-noise ratios (SNRs).
For example, on the Additive White Gaussian Noise (AWGN) channel with SNR 
smaller than 0 dB, the largest achievable SE by any 128-bit code at BLock
Error Rate (BLER) of $10^{-4}$ is less than half the channel capacity~\cite{bib:VucComMag19}.

Conventional ECC methods do not send any feedback information.
However, as the majority of contemporary communication systems have
two-way links, feedback channels are available in most cases.
It is therefore natural to seek improvements over conventional ECC
by making use of feedback.
A well-known information-theoretic 
result~\cite[Thm. 7.12.1 and Sec. 9.6]{bib:CoverThomas}
stipulates that feedback does not increase the capacity of
memoryless channels.
Even in the most favorable case --
instantaneous noiseless feedback -- the channel capacity remains unchanged.
This means that the largest rate at which reliable\footnote{Here,
	the word \emph{reliable} has the classical Shannon-theoretic meaning
	of \emph{arbitrarily low error rate}.} transmission is possible is
the same regardless of whether feedback is available or not.
Nevertheless, digital communication research has shown that usage of
feedback potentially brings significant advantages in terms
of improved reliability. Improving the reliability yields decreased BLER,
thereby producing increased spectral efficiency.

The improved reliability of feedback-based codes has been shown
for the first time in the pioneering work of Schalkwjik and
Kailath~\cite{bib:Sch66}.
The authors of~\cite{bib:Sch66} developed a
code and a corresponding iterative encoding procedure for channels with
noiseless feedback where the transmitter sends an information message in
an initial uncoded transmission followed by a number of subsequent
transmissions containing corrections for the initial transmission.
Correction signals are calculated based on the information fed back to
the transmitter through a feedback channel.
Transmission of corrections continues for a predefined number of times.
After the last transmission, the receiver delivers in its output an
accurate replica of the transmitted information.
The error performance analysis of the Schalkwjik-Kailath (SK) code
reveals that the error probability has a \emph{doubly-exponential
decay} with the codeword length $N$, as shown by the following equation~\cite{bib:Ben-Yishai-Shayevitz-17}:
\begin{equation}
	\label{eq:PwSK}
	P_w^{\rm (SK)} < \exp\left(-A\times 2^{2N(C-R)}\right).
\end{equation}
Here, $P_w^{\rm (SK)}$ is the probability that the SK decoder delivers
in its output a decoded message containing errors, $A$ is a
SNR-dependent term, $N$ is the codeword length,
$C$ is the channel capacity and $R$ is the code rate.
In comparison, the error probability of conventional ECC exhibits an
exponential decay with $N$, as shown by the following
equation~\cite{bib:Proakis}:
\begin{equation}
	\label{eq:PwECC}
	P_w^{\rm (ECC)} < \exp\left(-N E(R)\right).
\end{equation}
Here, $E(R)$ is the \emph{reliability function}~\cite{bib:Proakis},
also called the \emph{error exponent}. 
$E(R)$ is a monotonically decreasing function of $R$.
The faster decay with $N$ of the feedback codes' error probability predicted
by \eqref{eq:PwSK} is the main motivation behind our interest in this kind of
codes as it suggests that feedback codes potentially attain lower
error probabilities already with short codewords.

The SK code promises to achieve remarkable performance
compared to conventional ECC. 
However, its sensitivity to finite-precision numerical computations
precludes practical implementations.
To overcome the above shortcomings, several variations and
enhancements have been developed subsequently based on similar
principles.
A concise account of those solutions can be found
in~\cite[Ch. 17]{bib:ElGamalKim}.

One of the most recent developments aimed at overcoming the SK shortcomings
is the code architecture based on deep recurrent neural networks called \emph{Deepcode}~\cite{bib:Kim20}. 
Deepcode sends the uncoded information message in an initial
transmission and subsequently generates a sequence of parity-check
symbols based on the information message and on the past forward-channel
outputs.
Forward channel outputs are fed back to the encoder through a feedback
channel.
Neural network weights are obtained by jointly training the encoder
and decoder according to a conventional machine-learning procedure.
Performance evaluation in~\cite{bib:Kim20} show that Deepcode achieves
significantly lower error rates compared to conventional ECC.
However, there is a remaining gap between SE and the channel capacity.

Among the proposed solutions, the Compressed Error 
Cancellation (CEC) framework~\cite{bib:Ooi98} is one of the most
attracting approaches as it combines optimum reliability\footnote{
	''Optimum reliability'' means that CEC has the highest possible error exponent.}
with the proven ability to achieve a rate arbitrarily close to the channel capacity.
As a further benefit, CEC has low encoding/decoding complexity.
Similar as SK, CEC operates according to an iterative encoding
procedure.
The initial transmission delivers the information message.
A number of subsequent transmissions are performed in order to send
updates produced by the encoder based on the last transmission and on the channel
outputs obtained through feedback.
The updates are produced in two steps:
a \emph{source coding} step followed by a \emph{precoding} step. 
	Source coding produces a compressed word based on the
	last transmitted sequence and on the corresponding received sequence, 
	which is made available to the transmitter through the feedback channel.
	Precoding divides the source-coded bits into segments of
	fixed length and maps each segment to a real integer in a given range
 (e.g., $\{-M,\ldots, M\}$)
	through the cumulative distribution of the forward
		channel input sequence.
	In that way, the symbols in the precoded sequence are
	distributed according to the forward-channel capacity achieving distribution.
Performance evaluations in~\cite{bib:Ooi98} show that CEC performs 
reliable transmission of long information messages -- 1 Mbit length --
at a rate approaching the capacity of the AWGN channel with
0 dB SNR within a very small gap.
However, as it will be shown later in the performance evaluations,
CEC transmission of \emph{short} messages (here, ''\emph{short}''
refers to messages of length 100 bits or less) achieves significantly
smaller data rates compared to the channel capacity.

The Accumulative Iterative Code (AIC) of this paper 
builds upon the findings of~\cite{bib:Ooi98} to develop a new
solution that significantly reduces the rate-capacity gap for
short codeword transmission and, at the same time, is of practical
interest as its building blocks are often found in most radio
communication systems.
Major novelties of AIC are the following:
\begin{itemize}
\item {\bf Usage of Huffman source coding.}
	Huffman coding is preferred over other source coding methods because
	it produces the lowest expected codeword length in its output.
\item {\bf Usage of conventional modulations on the forward channel.}
	Conventional modulations are preferred to the precoding of~\cite{bib:Ooi98}
	as they are already present in most radio communication systems, they
	allow simple implementation and achieve rates close to channel capacity in
	wide SNR ranges.
\item {\bf Feedback based on quantized Log-Likelihood Ratios (LLRs).}
	Sending the LLRs of transmitted \emph{bits} on the feedback channel makes
	AIC independent of the modulation.
	Moreover, LLR quantization keeps the feedback data rate contained.
	The LLR quantization method here considered is optimal in the sense that,
	for given number of quantization levels, the quantization thresholds are
	determined so as to maximize the mutual information of modulator input and
	quantizer output.
\end{itemize}
Performance evaluations show that AIC achieves SE close to channel capacity
and arbitrarily low error rate at the same time, thereby providing significant
gains compared to conventional coding methods.
The rest of the paper is organized as follows: Sec.~\ref{sec:aic} describes AIC,
the structure of encoder an decoder, and the encoding and decoding procedures.
Sec.~\ref{sec:perfeval} shows the results of performance evaluation.
with conventional ECC.
Final observations and conclusions are given in Sec.~\ref{sec:conclusions}.

\begin{figure*}[!t]
	\centering
	\resizebox{\hsize}{!}{
		\includegraphics[clip=true,trim=4.3cm 7.0cm 3.8cm 4.7cm]{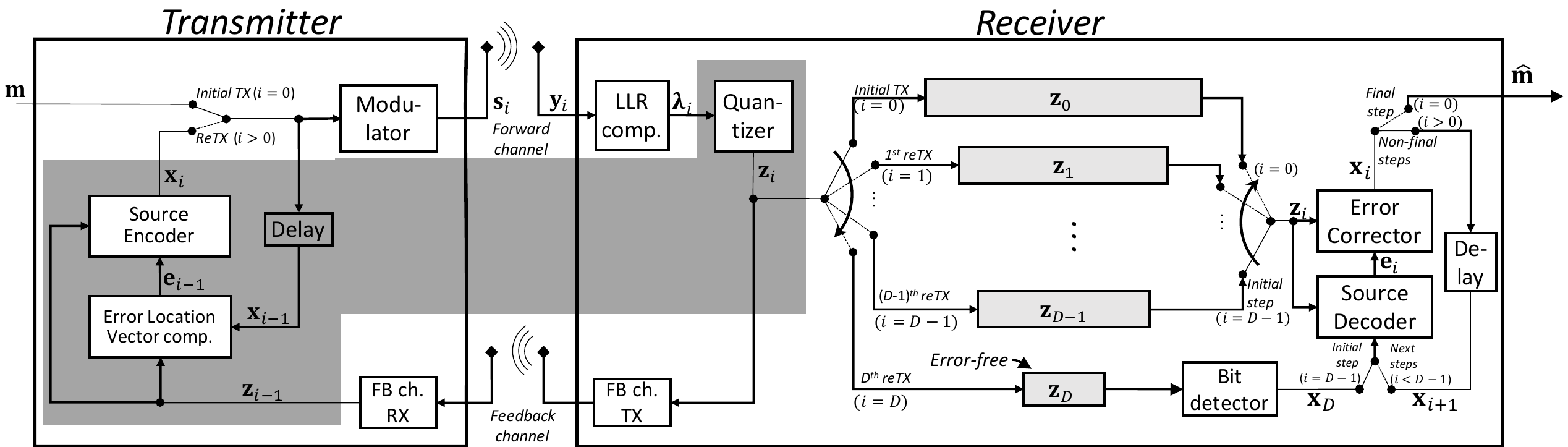}}
	\caption{AIC transmission system. The encoder is highlighted by a
		grey-shaded background.}
	\label{fig:AICencdec}
\end{figure*}



\section{The Accumulative Iterative Code}
\label{sec:aic}
Figure~\ref{fig:AICencdec} shows the block scheme of AIC
	transmission system where the encoder is highlighted by a grey-shaded background.
The AIC encoder is split between transmitter and receiver.
The encoder's transmitter component and receiver component interact with
each other by exchanging signals through the forward channel and the
feedback channel.
In each iteration, the encoder computes an \emph{error location vector} 
($\mathbf{e}_{i-1}$ in Figure~\ref{fig:AICencdec}) which contains 
the locations of errors in the last forward transmission.
Vector $\mathbf{e}_{i-1}$ is computed based
on the previously transmitted sequence $\mathbf{x}_{i-1}$ and on the
information obtained through the feedback channel
($\mathbf{z}_{i-1}$ in Figure~\ref{fig:AICencdec}).
The error location vector is source-coded so as to generate a
\emph{source-coded error location vector}
($\mathbf{x}_i$ in Figure~\ref{fig:AICencdec}).
The vector $\mathbf{x}_i$ is sent through the forward
channel so as to make error correction possible in the receiver.
However, as the received source-coded vector may again be
corrupted by errors, a further iteration is needed in order to inform
the decoder about error locations.
The feedback channel is assumed to be reliable -- the information transmitted
on the feedback channel is not corrupted by any errors.
In the receiver, Log-Likelihood Ratios (LLRs) of the received
	signals are quantized and stored in memory buffers.
	The Quantized LLRs (QLLRs) are also fed back to the transmitter through
	the feedback channel.

The AIC transmitter keeps sending error words on the forward
channel until an \emph{error-free forward transmission} occurs or until a
maximum number of iterations is reached.
When an error-free forward transmission occurs, the transmitter informs the
receiver that decoding can be performed.
Start of decoding can be signaled to the decoder, e.g., by sending an
acknowledgment (ACK) message through a control channel.
If a maximum number of iterations is reached and no error-free forward
transmissions occurred, the transmitter notifies the receiver (e.g., by
sending a negative acknowledgment -- NACK -- message through
the control channel) that transmission of the information message failed.

	After reception of the ACK message, the decoder
	starts the decoding process.
	In a first decoding step, the decoder determines the source-coded error
location vector that was transmitted in the last iteration
($\mathbf{x}_D$ in Figure~\ref{fig:AICencdec}).
In subsequent decoding iterations, the decoder determines the error location
vectors ($\mathbf{e}_{i}$ in Figure~\ref{fig:AICencdec}) by source-decoding the
vectors $\mathbf{x}_{i+1}$ and applies corresponding corrections.
In a final step, the decoder obtains the decoded message $\mathbf{\hat m}$.

The above encoding and decoding procedures are described in detail in
the subsections below.

\subsection{Encoding Procedure}
\label{subs:encproc}
The encoding procedure consists of a number of iterations, where in
each iteration the transmitter and receiver exchange signals through the
forward and feedback channels.
In the initial iteration, the transmitter sends a $K$-bit information message
$\mathbf{m} = (m_1, \ldots, m_K)$ on the forward channel; in subsequent iterations,
the transmitter sends source-coded error location vectors $\mathbf{x}_i, i=0,\ldots, D$,
where $\mathbf{x}_i = (x_{i,1}, \ldots, x_{i,N_i})$.
is the transmitted word in the $i^{\rm th}$ iteration and $N_i$ is its length.
For convenience, we define ${\bf x}_0 \triangleq {\bf m}$ and $N_0 \triangleq K$.

The message $\mathbf{m}$ and the source coded vectors $\mathbf{x}_i, i=0,\ldots, D$,
are transmitted on the forward channel using conventional modulations.
In the $i^{\rm th}$ iteration, the modulator generates a sequence
$\mathbf{s}_i = (s_{i, 1}, \ldots, s_{i,L_i})$, where $L_i = \left\lceil N_i/Q\right\rceil$ and $Q$ is the modulation order.
Each element of $\mathbf{s}_i$ is obtained by mapping a group of $Q$
consecutive bits (hereafter called a $Q$-tuple) of $\mathbf{x}_i$ to a
complex signal selected from a given set
$\Psi = \{\psi_1,\ldots,\psi_{2^Q}\}$ through a
one-to-one labeling map $\mu_Q:\{0,1\}^Q \to {\Psi}$.

The received sequence is obtained as follows:
\begin{equation}
	\mathbf{y}_i = \mathbf{s}_i +  {\pmb\nu}_i
	\label{eq:RXsignal}
\end{equation}
where $ {\pmb\nu}_i$ represents noise, interference and distortions
introduced by the forward channel.

Based on the received sequence $\mathbf{y}_i$, the encoder computes a sequence of LLRs
 of transmitted bits ${\pmb \lambda}_i = (\lambda_{i, 1}, \ldots, \lambda_{i, N_i})$
 as follows:
\begin{eqnarray}
	\lambda_{i,n} & = &  \log\frac{P(x_{i,n}=0\mid\mathbf{y}_i)}{P(x_{i,n}=1\mid\mathbf{y}_i)} \label{eq:LLR}\\ 
	& = & \log{\frac{\sum_{t \in {\Psi}_0^{p(n)}}{P(y_{i,q(n)} \mid s_{i,q(n)}=t)}}{\sum_{t \in {\Psi}_1^{p(n)}}{P(y_{i,q(n)} \mid s_{i,q(n)}=t)}}} \label{eq:LLRpart}
\end{eqnarray}
where $n=1,\ldots, N_i$, $P(\cal E)$ denotes the probability of event $\cal E$,
$q(n)$ is the position in the sequence $\mathbf{s}_{i}$ of the
modulation signal that carries $x_{i,n}$ and $p(n)$ is the position of
$x_{i,n}$ in the $Q$-tuple that produced that modulation signal.
${\Psi}_b^{p(n)}$ denotes the subset of the signals of $\Psi$
whose label has value $b$ in position $p(n)$, where $b\in \{0, 1\}$.

Each LLR in the above sequence is quantized so as to obtain a sequence of
QLLRs $\mathbf{z}_i = (z_{i, 1}, \ldots, x_{i, N_i})$ as follows:
\begin{equation}
z_{i,n} = Q_{\pmb\theta^*}(\lambda_{i,n}) \label{eq:QLLRs}
\end{equation}
where $\pmb\theta^* = (\theta_0^*,\ldots, \theta_{R-1}^*,\theta_R^* = +\infty)$
is a vector of non-negative \emph{quantization thresholds} arranged
in increasing order, i.e.:
\begin{equation}
	\theta_0^* < \theta_1^* < \ldots < \theta_R^*
\end{equation}
and $Q_{\pmb\theta^*}$ is a function that maps an arbitrary real value
$\lambda$ to an integer value in the set $\{0, \pm 1,...,\pm R\}$ as follows:
\begin{equation}
	\label{eq:quant}
	Q_{\pmb\theta^*}(\lambda) \triangleq \left\{\begin{array}{rcl}
		+r & \textrm{if} & \theta_{r-1}^* \leq \lambda < \theta_r^* \\
		0  & \textrm{if} & |\lambda| < \theta_0^* \\
		-r & \textrm{if} & -\theta_r^* < \lambda \leq -\theta_{r-1}^*
	\end{array}\right., r = 1,\ldots, R.
\end{equation}
The \emph{optimal} quantization thresholds vector -- denoted as 
$\pmb\theta^*$ -- is obtained by maximizing the amount of information
that is transferred from the modulator input to the quantizer output
through the forward channel.
Assuming that $X$ is a Random Variable (RV) representing the
modulator input and $Z$ is a RV representing the LLR quantizer
output, the vector $\pmb\theta^*$ is given by 
\begin{equation}
	\begin{split}
	\pmb\theta^* =
	\arg\max_{\pmb\theta}\sum_{u\in\{0,1\}}\sum_{v\in\{0, \pm1, \ldots,\pm R\}} p_{uv}(\pmb\theta) P_X(u)
		\log\frac{p_{uv}(\pmb\theta)}{P_Z(v)}
	\end{split}
\label{eq:thetaopt}
\end{equation}
where the probabilities $p_{uv}(\pmb\theta)$ are defined as follows:
\begin{align}\label{eq:condDMCprob}
	p_{uv}(\pmb\theta) = & P(Z = v \mid X = u;\pmb\theta), \nonumber \\ 
	& u \in \{0, 1\}, v \in \{0, \pm 1, \ldots, \pm R\}. 
\end{align}
Moreover, we assume uniform input distribution and thus
\begin{equation}
	P_X(u) \triangleq P(X=u) = 1/2, u\in\{0, 1\}
\end{equation}
and probabilities $P_Z(v)$ are obtained as follows:
\begin{equation}
	P_Z(v) \triangleq P(Z=v) = \sum_{u\in\{0,1\}}{p_{uv}(\pmb\theta)P_X(u)}.
\end{equation}
By taking into account quantization as mathematically modeled by equations
\eqref{eq:QLLRs} and \eqref{eq:quant},  equation
\eqref{eq:condDMCprob} yields:
\begin{equation}
	p_{uv}(\pmb\theta)\!=\!\left\{\!\begin{array}{ll}
		\!P(\lambda_{i,n}\!\in\![\theta_{v-1},\theta_{v})\!\mid\!x_{i,n} = u) & (v > 0) \\
\!P(\lambda_{i,n}\!\in\!(-\theta_{0}, \theta_{0})\!\mid\!x_{i,n} = u) & (v = 0) \\
\!P(\lambda_{i,n}\!\in\!(-\theta_{-v}, -\theta_{-v-1}]\!\mid\!x_{i,n} = u) & (v < 0)\end{array} \right.\label{eq:puvtheta1}
\end{equation}
The following example illustrates how \eqref{eq:puvtheta1} can be
computed in a practical case -- BPSK signals transmitted on the AWGN channel.
The procedure in the example can be straightforwardly extended to Gray-mapped
QPSK signals on AWGN, as the in-phase and quadrature components of QPSK
can be treated as independent BPSK-modulated signals.

{\bf\textit{Example.}} For BPSK signals corrupted by AWGN, \eqref{eq:LLRpart} can be simplified as follows:
\begin{equation}
	\lambda_{i,n} = \frac{2}{\sigma_\nu^2}y_{i,n}. \label{eq:LLRBPSK}
\end{equation}
By combining \eqref{eq:puvtheta1} and \eqref{eq:LLRBPSK}, after few algebraic transformations we obtain
\begin{equation}\label{eq:2ints}
	p_{uv}(\pmb\theta)\!=\!\left\{\!
	\begin{array}{lc}
		\!\int_{\frac{\sigma_{\pmb\nu}^2}{2}\theta_{|v|-1}-\phi(v)\mu_{\rm BPSK}(u)}^{\frac{\sigma_{\pmb\nu}^2}{2}\theta_{|v|}-\phi(v)\mu_{\rm BPSK}(u)}{f_{\pmb\nu}(y)}{\rm d}y & (v\neq 0) \\
	\!\int_{-\frac{\sigma_{\pmb\nu}^2}{2}\theta_{0}-\mu_{\rm BPSK}(u)}^{\frac{\sigma_{\pmb\nu}^2}{2}\theta_{0}-\mu_{\rm BPSK}(u)}{f_{\pmb\nu}(y)}{\rm d}y & (v = 0) \end{array}\right. 
\end{equation}
where $\mu_{\rm BPSK}(u)$ is the conventional BPSK mapping,
defined as follows
\begin{equation} \label{eq:BPSKmapping}
	\mu_{\rm BPSK}(u) \triangleq 1 - 2 u, \quad u \in \{0, 1\}.
\end{equation}
and $\phi(v)$ is the \emph{sign} function: $\phi(v) \triangleq v/|v|$.
The function $f_{\pmb\nu}(y)$ is the well-known Gaussian \emph{pdf} with
zero mean and variance $\sigma_{\pmb\nu}^2$:
\begin{equation}
	f_{\pmb\nu}(y) = \frac{1}{\sqrt{2 \pi} \sigma_{\pmb\nu}} e^{-y^2/(2\sigma_{\pmb\nu}^2)}.
\end{equation}

The above example showed how to compute the probabilities $p_{uv}(\pmb\theta)$ for a given modulation format based on the channel
noise \emph{pdf} $f_{\pmb\nu}(y)$.
For those cases where $f_{\pmb\nu}(y)$ is not known,
the probabilities \eqref{eq:condDMCprob} have to be calculated by
Monte Carlo simulation.

The QLLRs sequence $\mathbf{z}_i$ computed according to \eqref{eq:QLLRs}
is stored in receiver memory buffers and
sent back to the transmitter through the feedback channel.
Based on the QLLRs sequence ${\bf z}_i$ and on the corresponding transmitted
word $\mathbf{x}_i$, the transmitter determines the \emph{error location vector} 
$\mathbf{e}_i = (e_{i,1}, \ldots, e_{i,N_i})$ as follows:
\begin{equation}
	\mathbf{e}_i = \mathbf{\bar{x}}_{i} \oplus \mathbf{x}_{i} \label{eq:e_i}
\end{equation}
where $\oplus$ denotes bit-wise modulo-2 sum and the vector
\mbox{$\mathbf{\bar{x}}_{i}=(\bar x_{i,1}, \ldots, \bar x_{i,N_i})$} is
obtained as follows:
\begin{equation}
	\label{eq:x-bar}
	\bar{x}_{i,n} = \left\{\begin{array}{rcl}
		0 & \textrm{if} & {z}_{i,n} > 0 \\
		1 & \textrm{if} & {z}_{i,n} < 0
	\end{array}\right., n=1,\ldots,N_{i}.
\end{equation}
Thus, $\bar{x}_{i,n}$ is the most likely value of the transmitted bit
$x_{i,n}$ based on the corresponding QLLR $z_{i,n}$.

\begin{figure}
	\centering
	\resizebox{0.6\hsize}{!}{	
		\includegraphics[clip=true,trim=13.7cm 7.4cm 13.4cm 7.2cm]{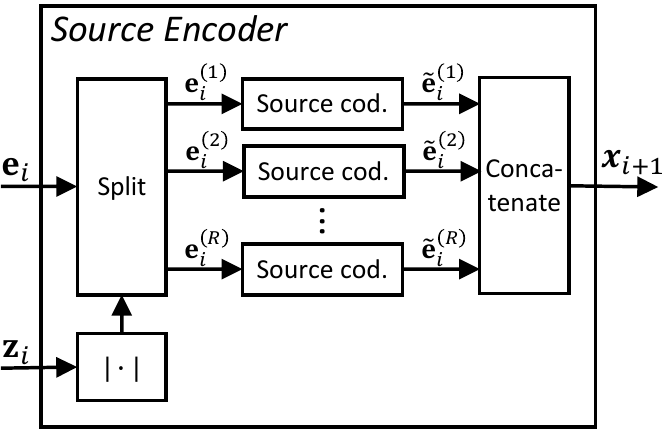}}
	\caption{Source encoder. $|\cdot|$ denotes absolute value.}
	\label{fig:SrcEnc}
\end{figure}
The vector $\mathbf{e}_i$ is source-coded so as to obtain a source-coded error location vector $\mathbf{x}_i$.
The source encoder is shown in Figure~\ref{fig:SrcEnc}.
Source coding of error location vectors is performed according to the
following steps:
\begin{enumerate}
	\item {\bf Splitting}. The error location vector $\mathbf{e}_i$ is split into $R$ subvectors $\mathbf{e}_i^{(r)}, r = 1, \ldots, R$, where $\mathbf{e}_i^{(r)}$ contains the elements of $\mathbf{e}_i$ whose corresponding QLLRs have absolute value $r$;
	
	\item {\bf Source coding}. Each subvector $\mathbf{e}_i^{(r)}$ is
	source-coded so as to obtain a source coded subvector
	$\mathbf{\tilde e}_i^{(r)}$.
	
	\item {\bf Concatenation}. The source coded sub-vectors
	$\mathbf{\tilde e}_i^{(r)}$ are concatenated so as to obtain a
	source coded error location vector  $\mathbf{x}_{i+1}$.
\end{enumerate}
When $\theta_0 > 0$, the source encoder further appends to $\mathbf{x}_{i+1}$
the bits of $\mathbf{x}_{i}$ whose corresponding QLLR is 0.

The source coding method we adopt is the Huffman method~\cite{bib:CoverThomas}.
For a given positive integer $H$, the Huffman encoder divides
its input vector $\mathbf{e}_i^{(r)}$ into\footnote{$\ell(\cdot)$
denotes vector length.}
$\lceil\ell(\mathbf{e}_i^{(r)})/H\rceil$ non-overlapping
segments $f_k$ of $H$ consecutive bits.
When $\ell(\mathbf{e}_i^{(r)})$ is not an integer multiple of $H$,
zeros are appended to $\mathbf{e}_i^{(r)}$ until the length becomes
multiple of $H$.
Assuming that the bits of $\mathbf{e}_i^{(r)}$ are statistically independent
and take value '1' with probability $\pi_r$,
the probability of a segment $f_k$  is given by:
\begin{equation}
	P(f_k) = \pi_r^{\pmb \omega(f_k)} (1-\pi_r)^{H-\omega(f_k)}
	\label{eq:encInAlphProb}	
\end{equation}
where $\omega(f_k)$ denotes the number of '1's in $f_k$.
Based on the above probabilities \eqref{eq:encInAlphProb},
the source encoder codebook is determined by the well known Huffman method~\cite{bib:CoverThomas}.
Once the codebook has been determined, the Huffman encoder
maps each segment $f_k$ to a corresponding codeword $\tilde{f}_k$ from the
codebook and produces a corresponding source-coded subvector $\mathbf{\tilde{e}}_i^{(r)}$
by concatenating the codewords $\tilde{f}_k$ as follows:
\begin{equation}
	\tilde {\mathbf{e}}_i^{(r)} = [\tilde{f}_1, \ldots, \tilde f_{\lceil\ell(\mathbf{e}_i^{(r)})/H\rceil}].
\end{equation}	
In the sequel, we will discuss how the probabilities $\pi_r$ of
\eqref{eq:encInAlphProb} are calculated.

We recall that a '1' in $\mathbf{e}_i^{(r)}$ indicates that a forward
transmission error occurred in a corresponding bit of $\mathbf{x}_i$.
According to the definition of LLR \eqref{eq:LLR},
a forward transmission error occurs if (i) the quantizer produces a negative QLLR
when the corresponding transmitted bit is '0', or (ii) the 
 quantizer produces a positive QLLR when the corresponding transmitted
 bit is '1'.
The above events have the following probabilities\footnote{In order to simplify notation,
	we omit to indicate the dependency of the channel transition
	probabilities on $\pmb\theta$ in the rest of this subsection.}:
\begin{eqnarray}
	\pi_{r,0} & = & P(Z < 0 \mid X = 0, |Z|=r)  \label{eq:pi_r1a}\\
	\pi_{r,1} & = & P(Z >0 \mid X = 1, |Z|=r). \label{eq:pi_r1b}
\end{eqnarray}
In general, the error probabilities \eqref{eq:pi_r1a} and \eqref{eq:pi_r1b} are not equal.
However, the source encoder treats both probabilities in the same way, that is,
it does not distinguish whether a bit error corresponds to a transmitted '1' or '0'.
Thus, it is necessary to make sure that there is a contained difference
between $\pi_{r,0}$ and $\pi_{r,1}$, that is:
\begin{equation}\label{eq:approxEq}
	\pi_{r,0} \cong \pi_{r,1}.
\end{equation}
Assuming that \eqref{eq:approxEq} holds, we obtain $\pi_r$ as follows:
\begin{equation}\label{eq:piRdef}
	\pi_r \triangleq \frac{\pi_{r,0} + \pi_{r,1}}{2}.
\end{equation}
In the following part of this subsection, we show how the probabilities
$\pi_{r,0}$ and $\pi_{r,1}$ are related to the conditional probabilities \eqref{eq:condDMCprob} and then, through that connection, we show that 
\eqref{eq:approxEq} holds with equality for BPSK and QPSK modulation signals
transmitted on the AWGN channel.
In Appendix~\ref{app:app} we show that \eqref{eq:approxEq} holds approximately
for higher-order modulations within their typical operating SNR ranges.

In order to show how $\pi_{r,0}$ and $\pi_{r,1}$ can be computed based \eqref{eq:condDMCprob}, we rewrite \eqref{eq:pi_r1a} and \eqref{eq:pi_r1b} as follows:
\begin{eqnarray}
	\pi_{r,0} & = & \frac{P(Z=-r \mid X = 0)}{P(|Z|=r)} \label{eq:pi_r2a}\\
	\pi_{r,1} & = & \frac{P(Z=r \mid X = 1)}{P(|Z|=r)}. \label{eq:pi_r2b}
\end{eqnarray}
We note that the numerators of \eqref{eq:pi_r2a} and \eqref{eq:pi_r2b} can be 
obtained from \eqref{eq:condDMCprob} by setting $u=0, v=-r$ and $u=1,v=r$.
Thus, by combining \eqref{eq:pi_r2a} and \eqref{eq:pi_r2b} with
\eqref{eq:condDMCprob} we obtain 
\begin{eqnarray}
	\pi_{r,0} & = & \frac{p_{0,-r}}{\rho_r} \label{eq:pi_r0}\\
	\pi_{r,1} & = & \frac{p_{1,r}}{\rho_r}. \label{eq:pi_r1}
\end{eqnarray}
where $\rho_r$ is the probability that $|Z| = r$.
$\rho_r$ can be obtained from the probabilities $p_{uv}$ of \eqref{eq:condDMCprob} as follows:
\begin{eqnarray}
	\rho_r & \triangleq & P(|Z|=r) \\
	& = & \sum_{u\in \{0, 1\}}{\sum_{v \in \{-r,r\}}{p_{uv} p_u}}.
\end{eqnarray}
For BPSK modulation signals transmitted on the AWGN channel,
$p_{0,-r}$ and $p_{1,r}$ can be derived from \eqref{eq:2ints} by setting $u=0, v=-r$ and $u=1, v=r$.
The following expression is obtained:
\begin{equation}
	p_{0,-r} = p_{1,r} = \int_{\frac{\sigma_{\pmb\nu}^2}{2}\theta_{r-1}+1}^{\frac{\sigma_{\pmb\nu}^2}{2}\theta_{r}+1}{f_{\pmb\nu}(y)}{\rm d}y.
\end{equation}
This equation, when combined with \eqref{eq:pi_r0} and \eqref{eq:pi_r1}, shows that \eqref{eq:approxEq} holds with equality.

\subsection{Decoding Procedure}
\label{subs:dec}
\begin{figure}
	\centering
	\resizebox{0.55\hsize}{!}{	
		\includegraphics[clip=true,trim=13.3cm 5.cm 11.8cm 6.3cm]{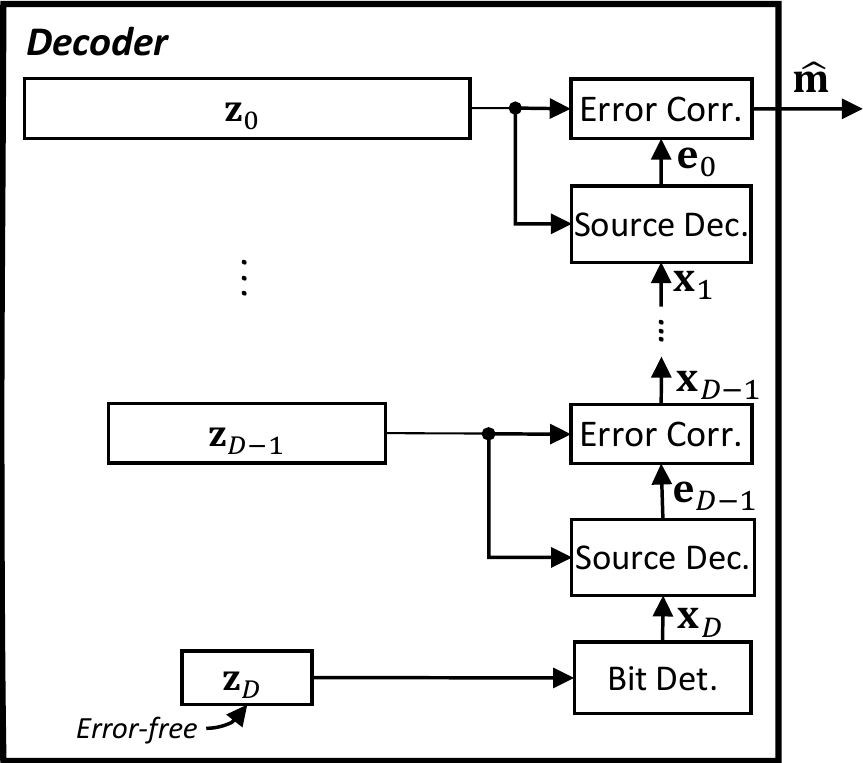}}
	\caption{AIC decoder.}
	\label{fig:AICdec}
\end{figure}

\begin{figure}[!t]
	\centering
	\resizebox{0.55\hsize}{!}{
		\includegraphics[clip=true,trim=13.4cm 7.4cm 13.5cm 7.2cm]{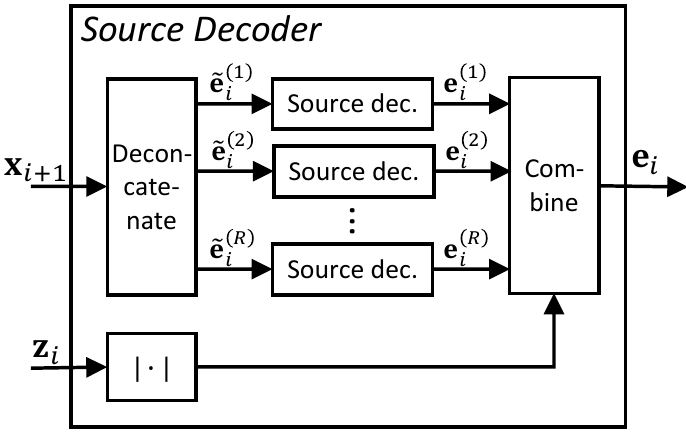}}
	\caption{Source decoder. $|\cdot|$ denotes absolute value.}
	\label{fig:SrcDec}
\end{figure}

The AIC decoder is shown in Figure~\ref{fig:AICdec}.
Decoding starts as soon as an error-free forward transmission occurs.
Let us assume that the $D^{\rm th }$ forward transmission has been
received free of errors.
The receiver starts the decoding process based on the QLLR vectors
${\bf z}_0, \ldots, {\bf z}_D$ which were stored in the receiver during
the encoding iterations (see Figure \ref{fig:AICencdec}, right side).
In a first decoding step, the decoder determines the $D^{\rm th}$
source-coded error location vector
${\bf x}_D = (x_{D,1}, \ldots, x_{D,N_D})$ based on the $D^{\rm th}$
QLLR word ${\bf z}_D$ as follows: 
\begin{equation}
	\label{eq:x-bar-i}
	x_{D,n} = \left\{\begin{array}{rcl}
		0 & \textrm{if} & {z}_{D,n} > 0 \\
		1 & \textrm{if} & {z}_{D,n} < 0
	\end{array}\right., n=1,\ldots,N_D.
\end{equation}
The above step is performed by the block labeled 
''\textsf{Bit detector}'' of Figure~\ref{fig:AICencdec}.
In a second step, the decoder determines the error location vector
$\mathbf{e}_{D-1}$ by source-decoding the word ${\bf x}_D$.
The source decoder is shown in Figure~\ref{fig:SrcDec}.
The decoder operates according to the following steps:
\begin{enumerate}
	\item {\bf Deconcatenation}. The source coded subvectors 
	$\mathbf{\tilde e}_{D-1}^{(r)}$ are obtained from the source
	coded error location vector  $\mathbf{x}_{D}$.
	
	\item {\bf Source decoding}. Each subvector $\mathbf{e}_{D-1}^{(r)}$
	is source-decoded so as to obtain a error location subvector
	$\mathbf{\tilde e}_{D-1}^{(r)}$;
	
	\item {\bf Combining}. The subvectors
	$\mathbf{e}_{D-1}^{(r)}, r = 1, \ldots, R$ are combined so as to 
	obtain the error location vector $\mathbf{e}_{D-1}$.
	Combining mirrors the source encoder's \emph{splitting} step,
	thereby restoring the original order of subvector elements in
	the error location vector $\mathbf{e}_{D-1}$.
\end{enumerate}

The decoder further performs error correction as follows:
\begin{equation}
	\mathbf{x}_{D-1} = \mathbf{\bar x}_{D-1} \oplus \mathbf{e}_{D-1}   
\end{equation}
where the vector $\mathbf{\bar x}_{D-1}$ is computed using~\eqref{eq:x-bar}
by setting $i=D-1$.

The above steps are repeated until the decoded information message ${\bf \hat m}$
is obtained by correcting the errors that corrupted the initial transmission 
as follows:
\begin{equation}
	{\bf \hat m} = \mathbf{\bar x}_{0} \oplus \mathbf{e}_{0}.
\end{equation}

\subsection{Average Codeword Length}
The AIC codeword is obtained as the concatenation of the information
message $\mathbf{m}$ and the source-coded error location words $\mathbf{x}_i, i=1,\ldots, D_{\rm MAX}$
in the following way:
\begin{equation}
	\mathbf{X} = [\mathbf{m}, \mathbf{x}_1, \ldots, \mathbf{x}_{D_{\rm MAX}}] .
\end{equation}
The length of $\mathbf{X}$, denoted by $N$, is a RV whose expected
value can be computed as follows:
\begin{equation}\label{eq:avgN}
	E[N] = E\left[\sum_{i=0}^{D_{\rm MAX}}N_i\right] = \sum_{i=0}^{D_{\rm MAX}}E\left[N_i\right]
\end{equation}
where $N_0$ is the message length, $N_i, i=1,\ldots,D_{\rm MAX},$ is the length of the $i^{\rm th}$
source-coded error location vector and $D_{\rm MAX}$ is the maximum number of
iterations.
Thanks to source coding, each source-coded error location vector has shorter
expected length compared to the length of the previous
source-coded error location vectors, that is:
\begin{equation} \label{eq:Niexp}
	E[N_i \mid N_{i-1}] < N_{i-1}.
\end{equation}
We prove \eqref{eq:Niexp} by noting that Huffman's
expected codeword length approaches the entropy rate of the
source that produces its input sequence as $H$ approaches infinity.
Thus, the source-coded error location vector length, denoted as
$\ell(\mathbf{\tilde e}_i^{(r)})$, has the following expected value:
\begin{equation}
	\lim_{H\to\infty}	E[\ell(\mathbf{\tilde e}_i^{(r)}) \mid \ell(\mathbf{e}_i^{(r)})] = {\cal H}_2(\pi_r) \ell(\mathbf{e}_i^{(r)}) \label{eq:expLen}
\end{equation}
where ${\cal H}_2(\pi_r) \triangleq -\pi_r \log_2 \pi_r - (1-\pi_r) \log_2 (1-\pi_r)$
is the \emph{binary entropy} function and $\pi_r$ is the probability
that a given bit of $\mathbf{e}_i^{(r)}$ is '1'.
Thus, when $H$ approaches infinity\footnote{The expected lengths derived
	in this section are all obtained for $H$ approaching 1.
	However, in order to simplify the exposition, the notation
	$\lim_{H\to\infty}$ will be omitted.}, the expected length of the
$i^{\rm th}$ source coded error location word $\mathbf{x}_{i}$ can be obtained as
follows:
\begin{eqnarray}
	E[N_{i} \mid N_{i-1}] &\stackrel{(a)}{=} & \sum_{r=0}^R{E[\ell(\mathbf{\tilde e}_{i-1}^{(r)}) \mid N_{i-1}]}\\
	& \stackrel{(b)}{=} & \sum_{r=0}^R{{\cal H}_2(\pi_r) E[\ell(\mathbf{e}_{i-1}^{(r)}) \mid N_{i-1}]} \\
	& = & \sum_{r=0}^R{{\cal H}_2(\pi_r) \rho_r N_{i-1}} \label{eq:33} \\
	& = & \alpha N_{i-1} \label{eq:alphaNi}
\end{eqnarray}
where, in the above chain of equations, (a) follows from the fact that
$\mathbf{x}_{i}$ is obtained by concatenation of compressed error subvectors
$\mathbf{\tilde e}_{i-1}^{(r)}$ and (b) follows from \eqref{eq:expLen}. 
Moreover, in \eqref{eq:alphaNi} we defined
\begin{equation}
	\alpha \triangleq \sum_{r=0}^R{{\cal H}_2(\pi_r) \rho_r}.
\end{equation}
Th prove \eqref{eq:Niexp}, it must be shown that $\alpha < 1$.
This inequality follows straightforwardly from the facts that $\rho_r$ is a probability distribution, therefore $\sum_{r=0}^R\rho_r=1$, and the binary entropy is ${\cal H}_2(\pi_r) \leq 1, \forall r=0,\ldots, R$.

By the law of total expectation, \eqref{eq:alphaNi} yields 
\begin{eqnarray} 
	E[N_{i}] &=& \sum_{k\in \mathbb{N}}E[N_{i} \mid N_{i-1}=k] P(N_{i-1} = k)\\
	& = & \sum_{k\in \mathbb{N}}\alpha k P(N_{i-1} = k) \\
	& = & \alpha E[N_{i-1}] \\
	& = & \alpha^{i} N_0 \label{eq:alphaN0}
\end{eqnarray}
where $N_0 = K$ is the message length.
Eq. \eqref{eq:alphaN0} shows that the expected lengths of the source-coded error
location vectors are exponentially decreasing according to a geometric
progression with common ratio $\alpha$.
By combining \eqref{eq:alphaN0} and \eqref{eq:avgN} we finally obtain
\begin{equation}\label{eq:41}
	E[N] =  \frac{1-\alpha^{D_{\rm MAX}+1}}{1-\alpha}K.
\end{equation}

The above derivations have been obtained in the limit of
$H$ approaching $\infty$. With finite $H$, the Huffman encoder
will produce codewords with larger average length, therefore
the right-hand side (RHS) of \eqref{eq:41} can be interpreted as a
lower bound to the average codeword length. In the following
section, we will use the RHS of \eqref{eq:41} to derive an upper bound
on the SE. We will show that the obtained upper bound is
tight in many cases, thereby proving that \eqref{eq:41} provides an
accurate prediction of the average codeword length obtained
using finite values of $H$ – predictions turn out to be accurate
even for rather small values of $H$, e.g., $H < 10$.

\section{Performance Evaluation}
\label{sec:perfeval}
In this section, we show the results of AIC performance evaluation.
In Subsec.~\ref{subsec:PerfEvalSE} we evaluate the AIC SE and compare
it with the SE achieved by conventional error correction codes through the 
Polyanskiy, Poor and Verd\'u (PPV) normal approximation~\cite{bib:PPV10}.
The comparison shows that AIC performs better than any conventional error correction code.
We also show that AIC performs better than the following feedback codes:
a feedback code with fixed codeword length -- Deepcode~\cite{bib:Kim20} --
and a feedback code with variable codeword length -- CEC~\cite{bib:Ooi98}.
In Subsec.~\ref{subsec:PerfEvalDistrib} we evaluate the cumulative
distributions of number of iterations and codeword length.
We compare the codeword length distribution of AIC with the distribution 
produced by NR HARQ and conclude that AIC's codeword length dispersion is
more contained than the dispersion of codeword lengths produced by NR HARQ.

\subsection{Spectral Efficiency}
\label{subsec:PerfEvalSE}
The performance of AIC is evaluated in terms of SE
vs. SNR of the received forward signal for given target BLER.
The SE is defined as follows: 
\begin{equation}
\label{eq:se}
{\rm SE} \triangleq \frac{K Q}{E[N]} (1 - {\rm BLER})\; \rm [bits/s/Hz]
\end{equation}
where $K$ is the message length and $Q$ is the modulation order.
The BLER is defined as follows:
\begin{equation}
	{\rm BLER} \triangleq P(\hat{\mathbf{m}} \neq \mathbf{m}).
\end{equation}

A simple analytical upper bound to the AIC spectral efficiency
can be obtained by combining \eqref{eq:41} and \eqref{eq:se}
with BLER = 0.
The resulting equation is the following:
\begin{equation}\label{eq:se_ub}
	SE_{\rm UB} \triangleq \frac{(1-\alpha) Q}{1 - \alpha^{D_{\rm MAX}+1}}.
\end{equation}

In the rest of this section, the average codeword length
$E[N]$ and BLER are evaluated by link-level simulation.
The corresponding SE is computed using \eqref{eq:se}.
For conventional codes, the evaluations are carried out
at a given target BLER ($10^{-4}$).
As for AIC, we let the encoder iterate
until error-free transmission occurs, thereby producing BLER = 0.
Such a comparison might be deemed unfair or inaccurate.
However, according to our observations, there is no noticeable 
difference between the SE of error-free AIC and the SE of AIC with
nonzero BLER as long as the BLER remains below $10^{-2}$.
Based on the above observation, we conclude that the comparison between
SE of error-free AIC and SE of conventional codes with BLER $=10^{-4}$
is accurate.

\begin{table}
	\centering
	\renewcommand{\arraystretch}{1.5}
	\begin{tabular}{c|c} 
		\hline
		\bf SNR [dB] & $\pmb\theta^*$ \\
		\hline \hline
		-2 & $(0, 1.42, \infty)$ \\
		0 & $(0, 1.72, \infty)$ \\
		2 & $(0, 2.07, \infty)$ \\
		4 & $(0, 2.47, \infty)$ \\
		6 & $(0, 2.92, \infty)$ \\
		\hline
	\end{tabular}
	\caption{Quantization thresholds obtained
		for QPSK modulation and $R=2$.}
	\label{tab:theta}
\end{table}

The forward channel is impaired by Additive White Gaussian Noise (AWGN) 
and by Quasi-Static Rayleigh fading (QSRF).
On the QSRF channel, each transmitted word
${\bf x}_i, i=0,\ldots, D$, is subject to a corresponding fading coefficient $h_i$,
as described by the following equation:
\begin{equation}
	\label{eq:qsrf}
	\mathbf{y}_i = h_i \mathbf{s}_i + {\pmb \nu}_i, \quad i= 0, \ldots, D
\end{equation}
where $h_i,i=0,\ldots, D$, are statistically independent 
fading RVs drawn from a Rayleigh distribution
with a suitably chosen mean so as to obtain unit 
average received signal energy.
The vector ${\pmb \nu}_i$ contains $N_i$ statistically independent
complex Gaussian noise signals with zero mean and variance $\sigma_{\pmb\nu}^2$.
The SNR for the $i^{\rm th}$ transmission is defined as follows:
\begin{equation}
	{\rm SNR}_i \triangleq \frac{h_i^2}{\sigma_{\pmb \nu}^2}.
\end{equation}
The performance evaluation results on the QSRF channel are
reported as BLER vs. average SNR, where the average SNR is defined as
follows:
\begin{equation}
	\overline{\mathrm{SNR}} = E[\mathrm{SNR}_i].
\end{equation}

The LLRs quantization thresholds are determined as
in \eqref{eq:thetaopt}, where the first threshold
is set as ${\pmb\theta}_0 = 0$.
Table~\ref{tab:theta} shows the values of optimal quantization thresholds
obtained with $R=2$ and SNRs between -2 dB and 6 dB.
The simulation parameters used for performance evaluation are summarized
in Table \ref{tab:EvalParams}.

\begin{table}
	\centering
	\renewcommand{\arraystretch}{1.5}
	\begin{tabular}{c|c}
		\hline
		\bf Parameter  & \bf Value \\
		\hline \hline
		Codeword length ($N$) [bits] & 128 \\ 
		Modulation & QPSK, 16QAM, 64QAM \\
		Modulation labeling & Gray \\
		Number of quantization thresholds ($R$) & $1, \ldots, 8$ \\
		Huffman dictionary size ($2^H$) & $2^{8}$ \\ 
		Maximum number of iterations ($D_{\rm MAX}$) & $\infty$ \\
		Target BLER & $10^{-4}$ \\
		\hline
	\end{tabular}
	\caption{Performance evaluation parameters.}
	\label{tab:EvalParams}
\end{table}

\begin{figure}
	\centering
	\includegraphics[clip=true,trim=3.8cm 2.cm 3.8cm 2.1cm, width=0.75\textwidth]{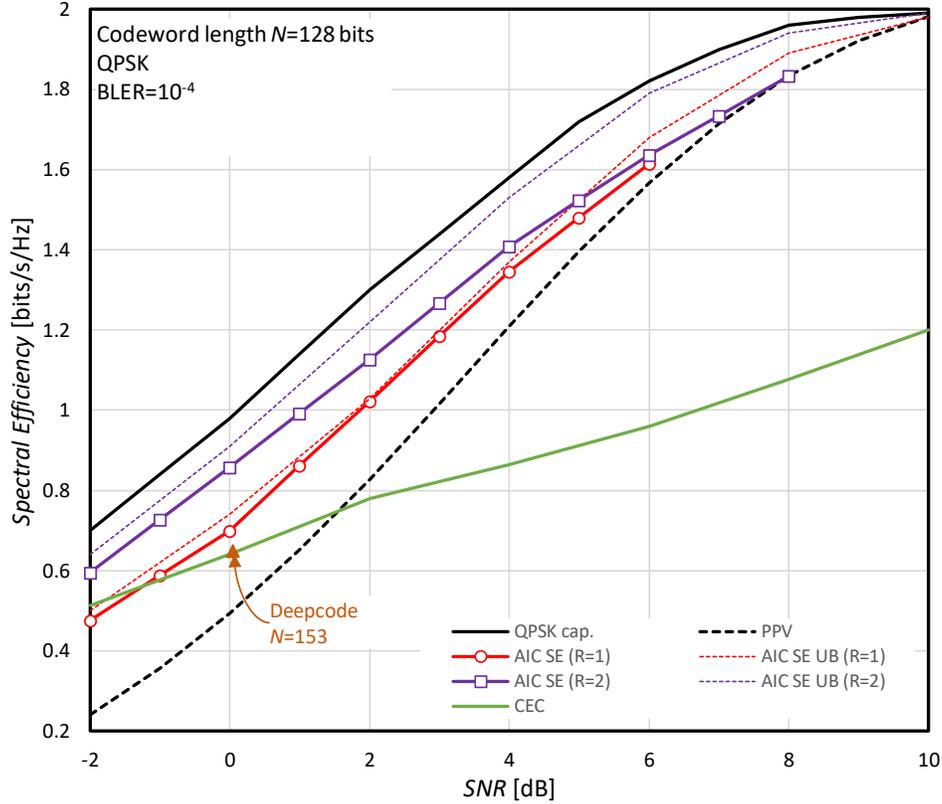}
	\caption{Spectral efficiency vs. SNR of QPSK-modulated AIC
		on AWGN. "AIC SE" denotes AIC spectral efficiency obtained
		by Monte Carlo simulation. "AIC SE UB" is the upper bound
		\eqref{eq:se_ub}.}
	\label{fig:SEvsSNR_QPSK}
\end{figure}

Figure~\ref{fig:SEvsSNR_QPSK} shows the SE vs. SNR performance of AIC
with QPSK modulation. 
The SE obtained with $R = 1$ is shown as a
solid red curve labeled "AIC SE (R=1)"; the corresponding SE
upper bound \eqref{eq:se_ub} is shown as a dashed red curve labeled
"AIC SE UB (R=1)".
The SE obtained with $R = 2$ is shown as a
solid purple curve labeled "AIC SE (R=2)"; the corresponding
SE upper bound \eqref{eq:se_ub} is shown as a dashed purple curve
labeled "AIC SE UB (R=2)".
The plot of Figure~\ref{fig:SEvsSNR_QPSK} also
shows the SE of the CEC method~\cite{bib:Ooi98}.
CEC has significantly
smaller SE compared to AIC, especially at high SNR. The
poor performance of CEC is mainly due to the Shannon-Fano
source coding of CEC – Shannon-Fano performance with short
codewords is worse than Huffman coding.
The Polyanskiy, Poor and Verd\'u (PPV) normal approximation~\cite{bib:PPV10}
is used to predict the performance that can be achieved by
state-of-art conventional codes.
To the best of the authors' knowledge, there is no conventional short
code that performs better than PPV, according to a summary of state-of-art
short codes' performance  in~\cite{bib:VucComMag19}.
It can be observed that AIC performs significantly better than PPV.
Even with a single quantization threshold ($R=1$), AIC provides SNR gains
larger than 1 dB in the range of SEs between 0.4 and 1 bits/s/Hz.
AIC with $R=2$ provides even larger gains as its performance approaches
the capacity of QPSK modulation up to within a small gap for
SNRs below 2 dB.
Values of $R>2$ do not provide significant gains compared to $R=2$.
Compared to Deepcode~\cite{bib:Kim20}, AIC with $R=2$ shows an SNR gain of
about 1.6 dB.
The upper bound~\eqref{eq:se_ub} provides an accurate prediction of
the real SE.
For $R=1$, the gap between upper bound and real SE is smaller than $5.5\%$
for all the evaluated SNRs in Figure \ref{fig:SEvsSNR_QPSK}.  
For $R=2$, the gap between upper bound and real SE is between $5.8\%$ and $9.4\%$
in the evaluated range of SNRs. The reason for the higher inaccuracy
observed with $R=2$ will be subject of investigation in future works.

\begin{figure}[!t]
	\centering
	\begin{subfigure}[t]{0.6\textwidth}
		\centering
		\includegraphics[clip=true,trim=2.2cm 3.6cm 2.1cm 3.5cm, width=\textwidth]{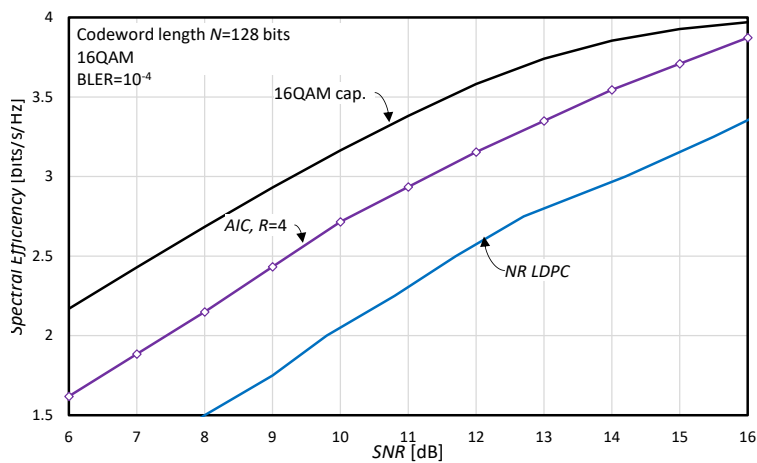}
		\caption{16QAM.}
		\label{fig:SEvsSNR_16QAM}
	\end{subfigure}
	\vfill 
	\begin{subfigure}[bt]{0.6\textwidth}
		\centering
		\includegraphics[clip=true,trim=2.2cm 3.6cm 2.1cm 3.5cm, width=\textwidth]{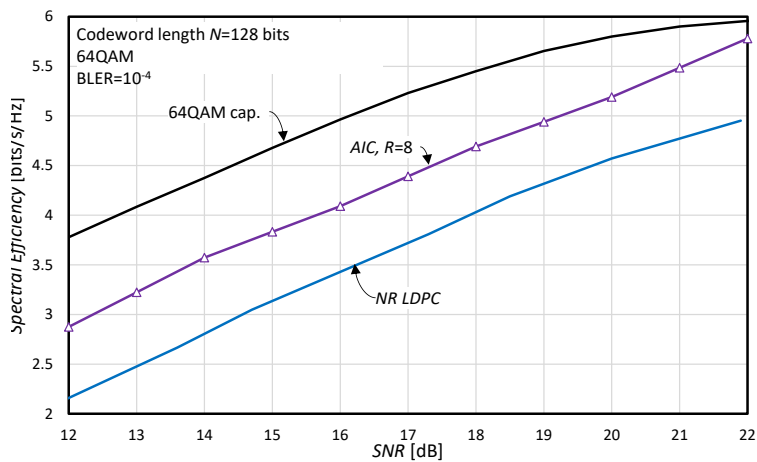}
		\caption{64QAM.}
		\label{fig:SEvsSNR_64QAM}
	\end{subfigure}
	\caption{Spectral efficiency vs. SNR of AIC on AWGN.}
	\label{fig:SEvsSNR}
\end{figure}

Figure~\ref{fig:SEvsSNR_16QAM} and Figure~\ref{fig:SEvsSNR_64QAM} show the
SE vs. SNR performance of AIC with 16QAM and 64QAM modulations.
With 16QAM and 64QAM, a larger number of quantization thresholds is needed
compared to QPSK, as each component of the QAM modulation signal uses 
more amplitude levels.
It has been found by numerical evaluation that $R=4$ is close to optimal
for 16QAM, while $R=8$ is close to optimal for 64QAM.
As there is no available theoretical result similar to PPV for 16QAM and
64QAM, we compare the AIC performance with
one of the best conventional channel codes known to-date -- NR LDPC codes.
The SNR gain of 16QAM/64QAM modulated AIC compared to the NR LDPC code with
same modulations is larger than 2 dB, whereas the AIC SE is 0.5 bits/s/Hz
larger than the SE of NR LDPC codes on the whole range of SNRs that we
evaluated.
These results show that AIC achieves arbitrarily high spectral efficiencies
by using conventional high-order modulations and provides remarkable gains
compared to conventional codes.

\begin{figure}[!t]
	\centering
	\resizebox{0.6\hsize}{!}{
		\includegraphics[clip=true,trim=2.2cm 3.6cm 2.2cm 3.5cm]{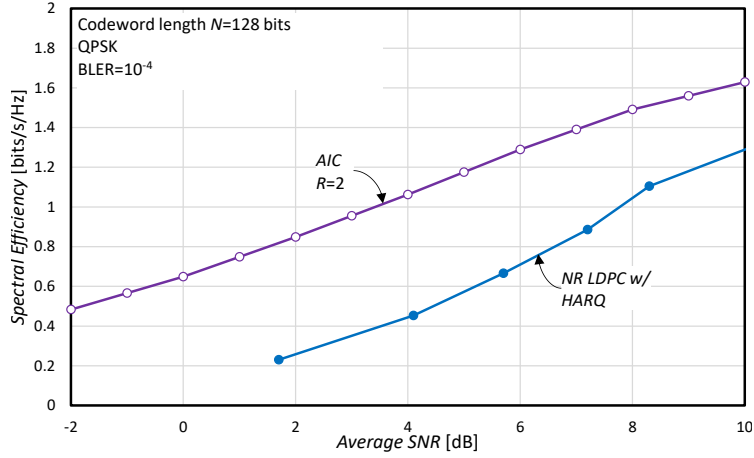}}
	\caption{Spectral efficiency of AIC with QPSK modulation on quasi-static
		Rayleigh fading channel.}
	\label{fig:SEvsSNR_QPSK_QSRF}
\end{figure}

Figure~\ref{fig:SEvsSNR_QPSK_QSRF} shows the spectral efficiency of AIC
with QPSK modulation on the QSRF channel described by \eqref{eq:qsrf}. 
The AIC SE is compared with the SE of the NR LDPC code with QPSK modulation
and Hybrid Automatic Repeat reQuest (HARQ)~\cite{bib:LinCostello, bib:NRr15-212}.
The reason for considering HARQ in the QSRF performance evaluation of LDPC codes
is that quasi-static fading combines detrimentally
with the channel dispersion~\cite{bib:PPV10}, that characterizes the AWGN
channel when used for transmission of short codewords, so as to make
reliable communication practically impossible at any reasonable SNR.
HARQ counteracts quasi-static fading by adaptively decreasing the code
rate based on the channel fading realization.
According to the evaluations shown in Figure~\ref{fig:SEvsSNR_QPSK_QSRF},
AIC shows a large SNR gain compared to NR LDPC codes with HARQ -- more than
4 dB for SE between 0.4 and 1.2 bits/s/Hz.

\subsection{Distribution of Number of Iterations and Codeword Length}
\label{subsec:PerfEvalDistrib}
\begin{figure}[!t]
	\centering
	\begin{subfigure}[t]{0.48\textwidth}
		\centering
		\includegraphics[clip=true,trim=2.7cm 2.1cm 3.cm 2.5cm, width=\textwidth]{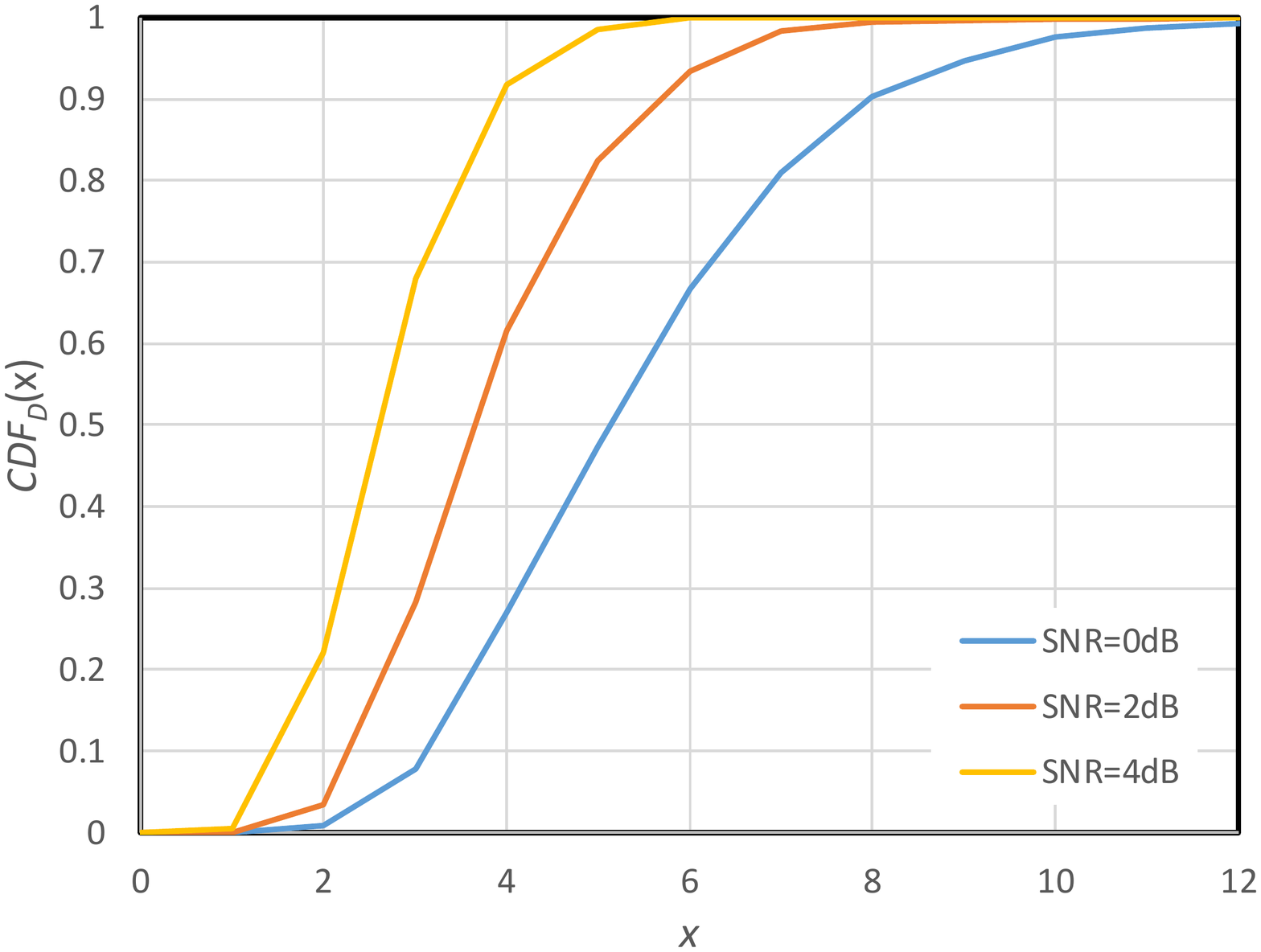}
		\caption{CDF of number of iterations \eqref{eq:cdf_D}.}
		\label{fig:Ntx_CDF}
	\end{subfigure}
	\hfill
	\begin{subfigure}[t]{0.48\textwidth}
		\centering
		\includegraphics[clip=true,trim=2.7cm 2.1cm 3.cm 2.5cm, width=\textwidth]{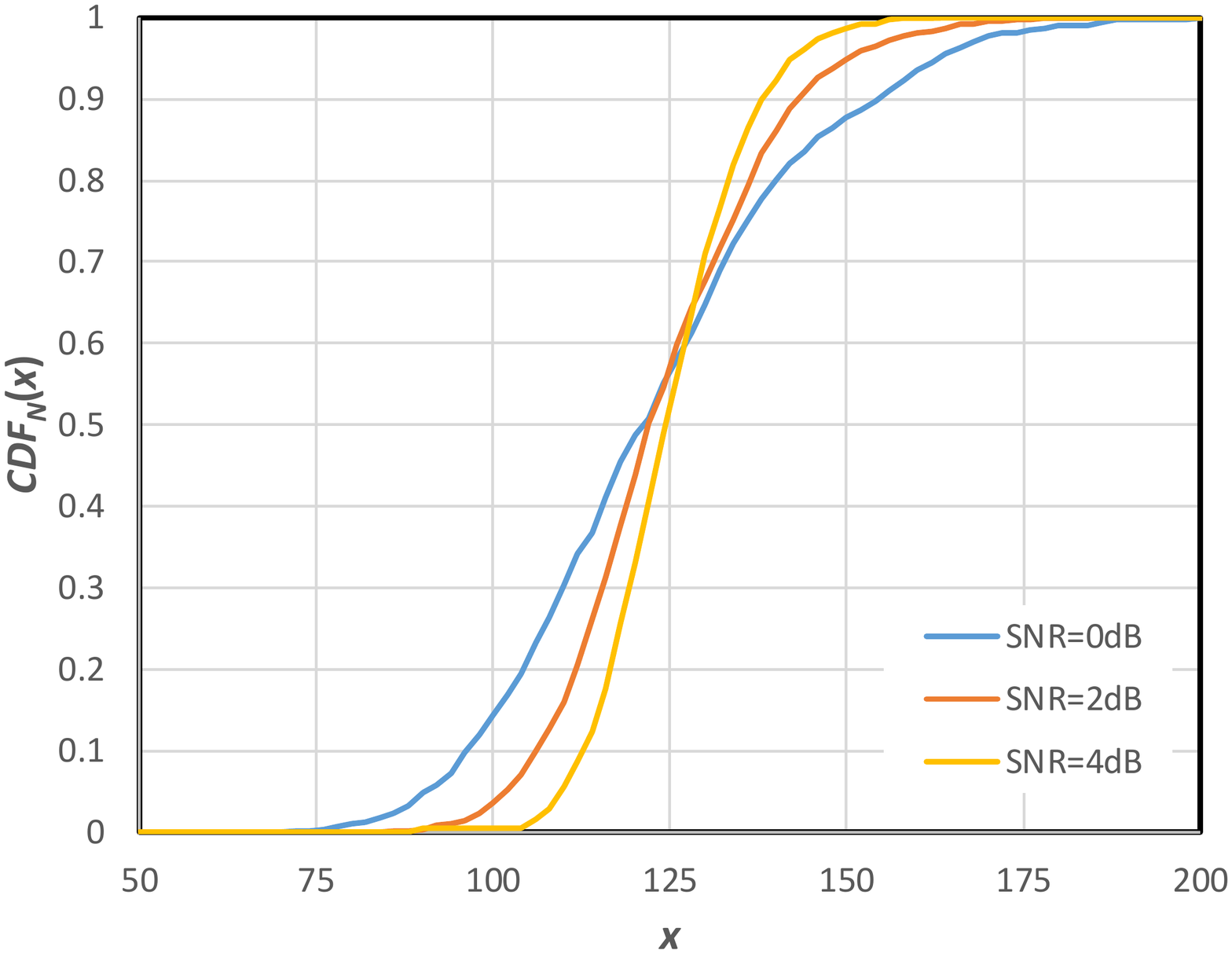}
		\caption{CDF of codeword length \eqref{eq:cdf_N}.}
		\label{fig:CWlen_CDF}
	\end{subfigure}
	\caption{Cumulative Distrbution Functions (CDFs) of
		(a) number of iterations $D$ and (b) codeword length $N$.}
	\label{fig:CDFs}
\end{figure}
The AIC encoding procedure performs a variable number of iterations and 
produces codewords of variable length -- 
number of iterations and the codeword lengths are unpredictable as
they ultimately depend on the forward channel noise and fading realizations.

Uncertainty on the number of iterations results in unpredictable transmission
delays -- a challenging situation for higher-layer protocols and for 
delay-sensitive applications.
In order to provide an assessment of AIC in terms of number of
iterations, we evaluate empirically the Cumulative Distribution
Function (CDF) of the number of iterations, which is defined as follows:
\begin{equation}\label{eq:cdf_D}
	CDF_D(x) = P(D \leq x)
\end{equation}
Figure~\ref{fig:Ntx_CDF} shows $CDF_D(x)$ for QPSK-modulated AIC
on AWGN channel with SNR=0 dB, 2 dB, and 4 dB.
The CDF curves are obtained by simulation with
1000 codewords.
It can be observed that for SNR=0 dB (blue curve), the number of
iterations needed to complete a message transmission takes
values between 2 and 13, where the average number of
iterations is approx. 5. With higher SNR, the number of
iterations is significantly smaller – 1 to 5 iterations for SNR=4
dB (yellow curve), where the average number of iterations is
approximately 2.5.

The number of iterations has a significant impact on
the transmission latency.
A contained number of iterations
produces shorter transmissions, thereby making the method
suitable for low-latency applications. 
AIC is a promising
candidate for low-latency applications thanks to its contained
\emph{average} number of iterations. 
However, AIC cannot guarantee
that all transmissions will be successfully completed in any
given number of iterations. 
A simple workaround to deal
with the above issue consists in stopping the transmission
after a maximum number of iterations $D_{\rm MAX}$, regardless of
whether there are remaining errors. If there are remaining
errors in the last iteration, the message transmission fails,
thereby producing BLER $> 0$. The relationship between BLER
and $D_{\rm MAX}$ is captured by the function $CDF_{D}(x)$ 
of \eqref{eq:cdf_D}:
for a given $D_{\rm MAX}$, $CDF_{D}(D_{\rm MAX})$ is the ratio of message
transmissions that require $D_{\rm MAX}$ or less iterations. Therefore,
for a given $D_{\rm MAX}$, BLER can be obtained as follows:
\begin{equation}
	BLER = 1 - {CDF}_{D}(D_{\rm MAX})
\end{equation}
It follows that, for a given target block error rate ${BLER}_{\rm T}$,
the maximum number of iterations needed to achieve that BLER is the following:
\begin{equation}
	D_{\rm MAX} = \min\{x\in \mathbb{N}: {CDF}_{D}(x) \geq 1 - {BLER}_{\rm T}\}
\end{equation}
or equivalently:
\begin{equation}\label{eq:D_MAX}
	D_{\rm MAX} = \lceil {CDF}_{D}^{-1}(1 - {BLER}_{\rm T})\rceil
\end{equation}
where ${CDF}_{D}^{-1}$ denotes the inverse function of ${CDF}_{D}$.
Eq. \eqref{eq:D_MAX} can be combined into \eqref{eq:41} and \eqref{eq:se_ub}
so as to provide expressions that capture the interplay between target BLER,
average codeword length and spectral efficiency.

Similar as the uncertainty on the number of iterations, the
uncertainty on the codeword length might be an issue for
transceiver design as the sizes of transmitter/receiver buffers
involved in encoding and decoding would have to be determined
based on worst-case situations.
In order to provide an
assessment of AIC in terms of codeword lengths, we evaluate
by simulation the codeword length CDF as follows:
\begin{equation}\label{eq:cdf_N}
	{CDF}_{N}(x) = P(N \leq x).
\end{equation}
Figure~\ref{fig:CWlen_CDF} shows a plot of ${CDF}_{N}(x)$ obtained
by simulation with 1000 codewords.
For SNR = 0 dB (blue curve), the codeword length $N$
takes values in a range between 72 bits and 206 bits.
As the SNR increases, the range shrinks -- for SNR = 4dB (yellow curve),
$N$ is distributed between 90 bits and 164 bits.
In Table \ref{tab:Ndisp}, the above values are summarized and a codeword
length dispersion value is computed as $N_{\rm max}/N_{\rm min}$ in the
rightmost column.
For comparison, NR HARQ is typically configured
to perform up to four transmissions, thereby producing
codewords whose length is up to four times the length of the
initial transmission, thus $N_{\rm max}/N_{\rm min}=4$.
Thus, the HARQ codeword lengths are typically spread over larger
intervals compared to AIC.
\begin{table}[hb]
	\centering
	\renewcommand{\arraystretch}{1.5}
	\begin{tabular}{c||c|c|c|c|c}
		\bf SNR & \bf $K$ & \bf $N_{\rm min}$ & \bf $N_{\rm max}$ & \bf $N_{\rm max}/N_{\rm min}$ & $N_{\rm max}/N_{\rm min}$ \\
		\bf [dB] & \bf [bits] & \bf [bits] & \bf [bits] & (AIC) & (NR HARQ)\\
		\hline
		0 & 54 & 72 & 206 & 2.86 & 4\\
		2 & 72 & 86 & 186 & 2.16 & 4\\
		4 & 90 & 90 & 164 & 1.82 & 4
	\end{tabular}
	\caption{Dispersion of codeword length $N$.}
	\label{tab:Ndisp}
\end{table}

It can be concluded that AIC's number of iterations and codeword length,
although being unpredictable, take values in rather contained intervals, 
thereby not posing significant challenges to transceiver design or
delay-sensitive applications.

\section{Conclusion}
\label{sec:conclusions}
A new error correction code for channels with feedback -- the
Accumulative Iterative Code -- has been described in this paper.
AIC encoder and decoder interact with each other by exchanging
signals through the forward and feedback channels.
The AIC encoder continues to perform iterations until an error-free
forward transmission occurs.

The new code achieves spectral efficiency close to channel capacity in
a wide range of SNRs even for transmission of short information messages
-- a situation where conventional ECC show a rather large gap between SE
and channel capacity.
In the same time, AIC provides arbitrarily low error rates, thereby being
suitable for applications demanding extremely high reliability.
Performance evaluations on the AWGN channel and quasi-static Rayleigh
fading show that AIC provides significant spectral efficiency and SNR
gains compared to conventional ECC methods.

Finally, it has been shown that the number of encoding iterations is
fairly small.
The codeword length, although unpredictable, takes
values in a contained interval, thereby not posing significant
challenges to transceiver design.

\appendices
\section{}
\label{app:app}
In Section~\ref{subs:encproc}, it has been shown that \eqref{eq:approxEq}
holds with equality for BPSK and Gray-mapped QPSK modulations.
We also claimed that \eqref{eq:approxEq} holds with the ``$\cong$'' sign for higher-order modulations within their typical operating SNR ranges. 
Here, we prove by numerical evaluations that the above statement is true.

Eq. \eqref{eq:approxEq} combined with \eqref{eq:pi_r0} and \eqref{eq:pi_r1} yields:
\begin{equation}\label{eq:approxPr}
	p_{0,-r}(\pmb\theta) \cong p_{1,r}(\pmb\theta).
\end{equation}
In order to prove \eqref{eq:approxPr}, we define a quadratic
dispersion – similar as the probabilistic concept of variance –
for the probabilities $p_{0,-r}(\pmb\theta)$ and $p_{1,r}(\pmb\theta)$
as follows:
\begin{equation}
	\delta_r \triangleq (p_{0,-r}(\pmb\theta) - E_r)^2 + (p_{1,r}(\pmb\theta) - E_r)^2
\end{equation}
where $E_r$ is the average of $p_{0,-r}(\pmb\theta)$ and $p_{1,r}(\pmb\theta)$,
and then we determine the maximum dispersion as follows:
\begin{equation}
	\Delta = \max_{r=0,\ldots,R}{\delta_r}
\end{equation}
The values of maximum dispersion obtained for 16QAM
and 64QAM within their typical SNR ranges are shown in
Table \ref{tab:Delta}.
It can be seen that the dispersion $\Delta$ remains very
contained for all SNR values.
We conclude that \eqref{eq:approxEq} holds for 16QAM and 64QAM within their typical SNR ranges of operation.

\begin{table}[t]
	\centering
	\renewcommand{\arraystretch}{1.5}
	\begin{tabular}{c||c|c}
		\bf SNR [dB] & \bf 16QAM & \bf 64QAM \\
		\hline
		6 & 5.72e-6 & -- \\
		8 & 4.55e-8 & -- \\
		10 & 1.60e-8 & -- \\
		12 & 9.09e-10 & 4.93e-7 \\
		14 & 1.22e-9 & 1.13e-8 \\
		16 & 7.74e-10 & 2.08e-9 \\
		18 & -- & 1.57e-9 \\
		20 & -- & 1.06e-9 \\
		22 & -- & 8.73e-11
	\end{tabular}
	\caption{Maximum dispersion $\Delta$ for 16QAM and 64QAM.}
	\label{tab:Delta}
\end{table}


\end{document}